\documentclass{article}
\usepackage{graphicx}
\usepackage{amsmath}
\usepackage{amssymb}

\begin{document}

\title{On the scaling properties of the total $\gamma^*\mathrm{p}$
cross section}

 \author{Miguel N. Mondragon, J.G. Contreras\thanks{
Corresponding author. E--mail: jgcn@mda.cinvestav.mx} \\ Departamento de
F\'{\i}sica Aplicada,\\ Centro de Investigaci\'on y Estudios Avanzados
del I.P.N., \\ 97310, M\'erida, Yucat\'an, M\'exico.}  \date{}
\maketitle
\begin{abstract}

We perform a detailed analysis on the scaling properties of the total
$\gamma^*\mathrm{p}$ cross section, $\sigma_{\gamma^*\mathrm{p}}$.  We
write the cross section as a product of two functions $W$ and $V$
representing, respectively, the dynamical degrees of freedom and the
contribution from the valence partons.  Analyzing data from HERA and
fixed target experiments we find that $V$ is nearly independent of
$Q^2$ and concentrated at large $x$, while $W$ carries all the
information on the $Q^2$ evolution of $\gamma^*\mathrm{p}$. We define
the reduced cross section $\tilde{\sigma}_{\gamma^*\mathrm{p}}
\equiv W=\sigma_{\gamma^*\mathrm{p}}/V$, and show that it is very close
to a generalized homogeneous function. This property gives rise to
geometric scaling not only for small $x$, but for all the current
measured kinematic plane.  As a consequence of our {\em Ansatz} we
also obtain a compact parameterization of
$\sigma_{\gamma^*\mathrm{p}}$ describing all data above $Q^2=1$
GeV$^2$.
\end{abstract}



\section{Introduction}
\label{intro}

It has been found that for low values of the Bjorken variable $x$,
$x\leq 0.01$, the total $\gamma^*\mathrm{p}$ cross section,
$\sigma_{\gamma^*\mathrm{p}}(x,Q^2)$, extracted from lepton--hadron
scattering presents the property of geometric
scaling~\cite{stasto,schildknecht}. This property permits to write the
cross section as a function of only one variable, $\tau$, called the
scaling variable, which is the product of two functions, one depending
only on $Q^2$ and the other only on $x$. It has been suggested that, for
$\sigma_{\gamma^*\mathrm{p}}$, $\tau$ is
given by $Q^2/Q^2_s$ with $Q^2_s=Q^2_s(x)$ known as the saturation
scale.  Geometric scaling has also been observed in $eA$
reactions~\cite{freund}, inclusive charm production~\cite{goncalves}
and nucleus--nucleus collisions~\cite{armesto}.

The observation that $\sigma_{\gamma^*\mathrm{p}}$ grows quite rapidly
at low $x$ and that this behavior can not continue indefinitely
without violating the unitarity of the cross section led to the
proposal of nonlinear QCD equations containing
saturation~\cite{gribov,levin,mueller,mclerran,balitsky,kovchegov}. 
One of the features of this type of
equations is the introduction of a scale, $Q^2_s$, to signal the onset
of saturation effects. Much of the excitement and advance in the
understanding of perturbative QCD at low $x$ in recent years comes
from the discovery that some saturation equations imply geometric
scaling at the saturation scale~\cite{bartels,tuchin,iancu,munier}.

Much recent theoretical work has been devoted to find both, the region
where geometric scaling is valid and the functional form of the
violations to geometric scaling above the saturation scale. Studies
based on the BFKL equation~\cite{bfkl} supplemented with specific
boundary conditions have found that there is a window of phase space
above $Q^2_s$ and below a $Q^2_{\mathrm{max}}$ on which geometric
scaling is valid. For current accessible energies,
$Q^2_{\mathrm{max}}$ is of the order of 100
GeV$^2$~\cite{kwiecinski,itakura,triantafyllopuolos}.  More recently,
there have been indications that, in a more general nonlinear
equation, geometric scaling is strongly violated~\cite{mueller2}. 

As the concept of geometric scaling has been linked to saturation,
none of these investigations expect geometric scaling to be valid 
at large $x$ where the density of partons is very small.

Here we study in detail the scaling properties of the total
$\gamma^*\mathrm{p}$ cross section and find that geometric scaling is
related to the fact that at small $x$, $\sigma_{\gamma^*\mathrm{p}}$
is very close to a homogeneous function, specifically a power law in
both, $x$ and $Q^2$. We show that it is possible to define a reduced
cross section, called $\tilde{\sigma}_{\gamma^*\mathrm{p}}$ in the
following, which isolates this power law behavior not only for the small,
but also for the large $x$ region and thus shows geometric scaling in the
complete kinematic plane.

This document is organized as follows.  In the next section, we show
that it is possible to isolate the power law behavior in $x$ of
$\sigma_{\gamma^*\mathrm{p}}$ for all values of $Q^2$ and define the
reduced cross section $\tilde{\sigma}_{\gamma^*\mathrm{p}}$. In
Section \ref{s:behavior} we study the behavior of
$\tilde{\sigma}_{\gamma^*\mathrm{p}}$ and show that it is very close
to a generalized homogeneous function. In Section \ref{s:discussion}
we discuss the implications of our findings regarding saturation and
geometric scaling. We also present a compact parametrization of
$\sigma_{\gamma^*\mathrm{p}}$ which describes all data above $Q^2=1$
GeV$^2$.  Finally, in Section \ref{s:summary} we briefly summarize our
findings and present our conclusions.

\section{Analysis of the $\mathbf{x}$ and $\mathbf{Q^2}$ dependence 
of $\mathbf{\sigma_{\gamma^*\mathrm{p}}}$}

First, we turn to the behavior of $\sigma_{\gamma^*\mathrm{p}}$ at
small $x$. It is know, that the experimental data at small $x$ can be
described at each value of $Q^2$ by a power law in $x$~\cite{h1}.
From the point of view of theory, this behavior is expected, for $Q^2$
big enough to justify the use of pQCD, from both, the DLLA
approximation \cite{derujula} if the starting $Q^2$ value for the QCD
evolution is taken sufficiently small (see for example \cite{kwie}),
and from the BFKL \cite{bfkl} evolution.

In fact, this behavior is also seen in studies of geometric scaling
above the saturation scale. For concreteness we use the saturation
scale $Q^2_s(x)\sim x^{-\lambda_{\mathrm{GBW}}}$, with
$\lambda_{\mathrm{GBW}} = 0.288$, as defined by K. Golec-Biernat and
M. W\"usthoff~\cite{gbw}. Sta\'sto {\em et al.}~\cite{stasto} found
that, for $\tau=Q^2/Q^2_s(x)>1$,
$\sigma_{\gamma^*\mathrm{p}}(\tau)\sim 1/\tau$ suggesting a power law
behavior of $\sigma_{\gamma^*\mathrm{p}}$ as a function of $x$ for
constant values of $Q^2$.

In summary, it is expected from theory and phenomenology, and confirmed
by experiment that, at small $x$, $\sigma_{\gamma^*\mathrm{p}}$ behaves
as a power law for fixed values of $Q^2$. Following these results, 
we propose to write the total $\gamma^*\mathrm{p}$ cross section at each
$Q^2$ value as
the product of two functions $W$ and $V$:

\begin{equation}
\sigma_{\gamma^*\mathrm{p}}=WV
\label{eq_wv}
\end{equation}
where 

\begin{equation}
W=N \bar{x}^{-\lambda},
\label{eq_w}
\end{equation}
and we use $\bar{x}$ instead of
$x$--Bjorken~\cite{gbw}. Both are related through

\begin{equation}
\bar{x}=x\left(1+\frac{4m^2_f}{Q^2}\right),
\end{equation}
with $m_f$ = 140 MeV. Note that $x$ and $\bar{x}$ are substantially
different only at very low $Q^2$. Note also that in equation
(\ref{eq_w}) both, the normalization, $N$, and the exponent of the power
law, $\lambda$, may be different, for different values of $Q^2$; i.e.,
$N=N(Q^2)$ and $\lambda=\lambda(Q^2)$.

As we are interested in isolating the scaling behavior of the cross
section we require from $V$ to be approximately constant at small $x$,
so that it does not alter the physics embodied in $W$, and that it
describes the cross section at large $x$. It turns out that these
requirements are very closed to those expected from a valence
distribution, so we tried the following functional form inspired in the
work of \cite{edin1,edin2}:

\begin{equation}
V=
\exp\left(-\frac{(\bar{x}-x_0)^2}{4\sigma^2}\right)
\mathrm{erf}\left(\frac{1-(x-x_0)}{2\sigma}\right),
\label{eq_v}
\end{equation}
where the Gaussian represents the valence distribution of
the proton and the error function takes care of enforcing the
different kinematic constrains.

\begin{figure}
\begin{center}
\includegraphics*[width=.98\textwidth]{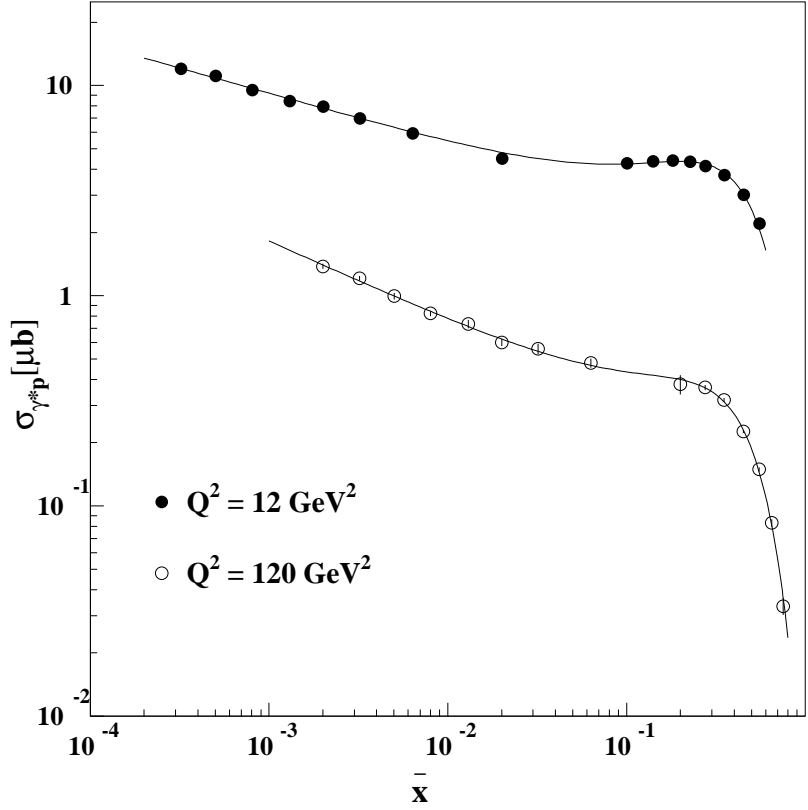}
\end{center}
\caption{The total $\gamma^*\mathrm{p}$ cross section is shown as a
function of $\bar{x}$ for $Q^2$=12 GeV$^2$ and $Q^2$=120 GeV$^2$ . The
points are experimental data from fixed target and HERA experiments
and the line is a fit to the form of equations
(\ref{eq_wv}--\ref{eq_v}). A power law is clearly seen at low values
of $\bar{x}$, while the structure at high $\bar{x}$ corresponds to a
Gaussian--like distribution.}
\label{fig1}
\end{figure}

We proceed now to test the {\em Ansatz} embodied in equations
(\ref{eq_wv}--\ref{eq_v}). We use data from fixed target \cite{bcdms}
and HERA \cite{h1data,zeus} experiments. We chose $Q^2$ values such
that there are measurements from both, fixed target and HERA
experiments, either at the same or at very similar $Q^2$. 
Normally, these $Q^2$ values are slightly different. We correct the fixed
target data to the HERA values using the H1 PDF 2000 fit
\cite{h1data}.  In most cases the correction factors are at the
per mil level with a few cases at the one and two percent level. We
then fit the data to the functional form of equation (\ref{eq_wv}) for
each $Q^2$ value. As an example, Figure \ref{fig1} shows data from
HERA and fixed target experiments at $Q^2$=12 GeV$^2$ and $Q^2$=120
GeV$^2$ along with the result of the fit.

%%%%%%%%%%%%%%%%%%%%%%%%%%%%%%%%%%%%%%%%%%%%%%%%%%%%%%%%
%%%%%%%%%%%%%%%%%%%%%%%%%%%%%%%%%%%%%%%%%%%%%%%%%%%%%%%%

The data in Figure \ref{fig1} is very well described by equation
(\ref{eq_wv}). Furthermore, the same can be said at each value of
$Q^2$ where there are measured points from both, fixed target and HERA
experiments. Remarkably, we find that $x_0$ and $\sigma^2$ from
equation (\ref{eq_v}) do not depend on $Q^2$. They take the values
$x_0 = 0.27$ and $\sigma^2=0.036$.

The fact that $V$ is independent of $Q^2$ implies that all the QCD
evolution in $Q^2$ of the cross section is entirely contained in $W$.
Thus, we define, in the complete kinematic plane, the reduced cross
section $\tilde{\sigma}_{\gamma^*\mathrm{p}}$ as

\begin{equation}
\tilde{\sigma}_{\gamma^*\mathrm{p}}
\equiv W(x,Q^2)=\sigma_{\gamma^*\mathrm{p}}/V(x).
\end{equation}

\section{Analysis of the behavior of 
$\mathbf{\tilde{\sigma}_{\gamma^*\mathrm{p}}}$}
\label{s:behavior}

We turn now to study the reduced cross section. Specifically, we study
the dependence on $Q^2$ of both $N$ and $\lambda$.  We use all HERA
data~\cite{h1data,zeus} to fit $\tilde{\sigma}_{\gamma^*\mathrm{p}}$
for fixed values of $Q^2$. We do not use fixed target data at this
stage of the analysis, because those data points are concentrated at
high $x$ and thus, they do not have a lever arm long enough to
determine accurately the power law parameters. Their influence has
already been taken into account through  $V(x)$.

\begin{figure}
\begin{center}
\includegraphics*[width=.98\textwidth]{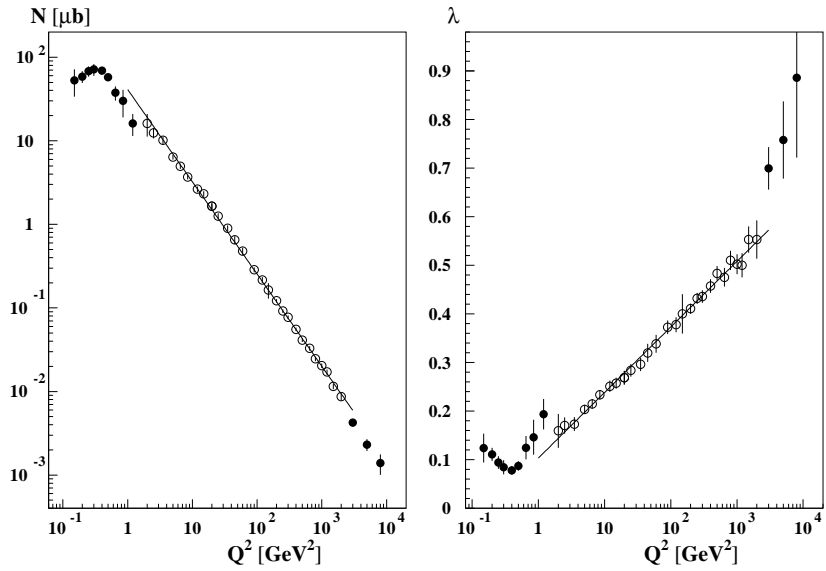}
\end{center}
\caption{The $Q^2$ dependence of the normalization $N$ and the
exponent $\lambda$ of $\tilde{\sigma}_{\gamma^*\mathrm{p}}
=N\bar{x}^{-\lambda}$, extracted from fits of HERA data to the reduced
cross section $\tilde{\sigma}_{\gamma^*\mathrm{p}}$ at fixed values of
$Q^2$. The solid lines are fits to equations (\ref{eq_l}) and
(\ref{eq_n}) in the intermediate $Q^2$ range given by the empty
bullets in the figure. }
\label{fig3}
\end{figure}

We use 40 different experimental $Q^2$ values, ranging from 0.15
to 8000 GeV$^2$, with enough data points in $x$ to perform the fit.
The average number of points for each fit was 8, ranging from 5 to
12. At each value of $Q^2$, equation (\ref{eq_w}) provides an
excellent description of data.  The results for $N$ and $\lambda$ for
each individual fit are quite precise and provide a clear picture of
their $Q^2$ dependence as shown in Figure~\ref{fig3}.

%%%%%%%%%%%%%%%%%%%%%%%%%%%%%%%%%%%%%%%%%%%%%%%%%%%%%%%%
%%%%%%%%%%%%%%%%%%%%%%%%%%%%%%%%%%%%%%%%%%%%%%%%%%%%%%%%

We observe a dramatic change in the behavior of both functions,
$N(Q^2)$ and $\lambda(Q^2)$, when the virtuality of the photon
approaches from above the region below 1 GeV$^2$. Here $\lambda$ is,
as expected, very similar to the one found with the
Donnachie--Landshoff parametrization \cite{dl}; while the normalization
appears to saturate. Furthermore, the value of $N$ and $\lambda$ for
the  $Q^2$ data  below 1 GeV$^2$ are almost constant in comparison to
the steep dependence of these functions above 1 GeV$^2$. 

Above $Q^2\approx 1$ GeV$^2$, $\lambda(Q^2)$ can be described with the
following functional form

\begin{equation}
\lambda(Q^2)=\alpha\log_{10}(Q^2/\Lambda^2),
\label{eq_l}
\end{equation} 
while $N(Q^2)$ behaves as the power law

\begin{equation}
N(Q^2)=\beta\left(\frac{Q^2}{Q^2_0}\right)^{-(1+\epsilon)},
\label{eq_n}
\end{equation}
where $Q^2_0$ is taken as 1 GeV$^2$.

A fit in the intermediate $Q^2$ region to the data plotted in Figure
\ref{fig3} yields $\beta=41.1\pm 1.5$ $\mu$b,
$\epsilon=0.104\pm0.007$, $\alpha=0.135\pm0.003$ and
$\Lambda^2=0.17\pm 0.03$ GeV$^2$. The quality of the fits is
$\chi^2/dof=0.66$ and $\chi^2/dof=0.33$ for $N(Q^2)$ and
$\lambda(Q^2)$ respectively. Note that the points at the largest $Q^2$
were not taken into account, because they have big fluctuations due to
the limited statistics of data.

In summary, for $Q^2$ values below 1 GeV$^2$ the $Q^2$ dependence of
$W$ is very soft and the function depends mainly on one variable: $x$.
Above 1 GeV$^2$ $W$ is very well described by the following functional
form:

\begin{equation}
W(\bar{x},Q^2) = 
\beta\left(\frac{Q^2}{Q^2_0}\right)^{-(1+\epsilon)}
\bar{x}^{-\alpha\log_{10}(Q^2/\Lambda^2)}. 
\label{eq_wxq2}
\end{equation}

%%%%%%%%%%%%%%%%%%%%%%%%%%%%%%%%%%%%%%%%%%%%%%%%%%%%%%%%
%%%%%%%%%%%%%%%%%%%%%%%%%%%%%%%%%%%%%%%%%%%%%%%%%%%%%%%%
\section{Discussion}
\label{s:discussion}

{\bf Scaling.} The $Q^2$ dependence of $\lambda$ is only
logarithmic. Consider first the case where the exponent of $x$ is a
constant $\lambda = \lambda_0$. Naming $W_0$ the function $W$ with
$\lambda_0$, equation (\ref{eq_wxq2}) would be of the form

\begin{equation}
W_0(x,Q^2) = k (Q^2)^{-(1+\epsilon)}x^{-\lambda_0},
\end{equation}
where $k$ is just the normalization and where we have reverted to $x$
instead of $\bar{x}$ for simplicity, because both are
equal in this kinematic domain.

In this case $W_0$ is a generalized homogeneous function, which, as can
easily be demonstrated, implies that
for all $t$ real and bigger than zero the following equation is
valid:

\begin{equation}
W_0(t^{-1/\lambda_0}x,t^{1/(1+\epsilon)}Q^2)=W_0(x,Q^2).
\end{equation}

In particular, it is also valid for $t=x^{\lambda_0}$:
\begin{equation}
W_0|_{t=x^{\lambda_0}}(1,x^{\lambda_0/(1+\epsilon)}Q^2)
\equiv W_0(\tau_0)=W_0(x,Q^2);
\end{equation}
i.e., $W_0$ exhibits {\em exact}  scaling with the scaling variable
given by $\tau_0=x^{\lambda_0/(1+\epsilon)}Q^2$.

Now, we turn to the real case where $\lambda$ depends on $Q^2$. Using
the same $t=x^{\lambda_0}$ one obtains

\begin{equation}
W(\tau_0)=W(x,Q^2)x^{-\delta},
\label{eq_ngs}
\end{equation}
where

\begin{equation}
\delta \equiv \lambda_0-\alpha\log_{10}(Q^2/\Lambda^2).
\label{eq_delta}
\end{equation}

In this case the scaling is broken when $\delta$, the exponent of $x$
in the RHS of equation (\ref{eq_ngs}) is different from zero. With the
help of Figure \ref{fig3}, note that for values of $Q^2$ between 1
GeV$^2$ and the upper limit of $Q^2$ used in \cite{stasto}, $Q^2 =
450$ GeV $^2$ (implicit for $x<0.01$), the average value of
$\lambda(Q^2)$ is close to the value of $\lambda_{\mathrm{GBW}}$.
Therefore, for $\lambda_0 = \lambda_{\mathrm{GBW}}$ the exponent of
$x$ in the RHS of equation (\ref{eq_ngs}) takes its smallest
values. Furthermore, one can check that for any given $Q^2$ in $1 <
Q^2 < 450$ GeV$^2$, the range covered in $x$ ($x< 0.01$) is limited in the
current measured kinematic plane to an average of one order of
magnitude. That means that for the given $Q^2$ the departure from
scaling for all $x$ points is similar and numerically small. As in
the small $x$ region $\tilde{\sigma}_{\gamma^*\mathrm{p}}\sim
\sigma_{\gamma^*\mathrm{p}}$, this explains the approximated geometric
scaling observed in ~\cite{stasto}.

This argument is valid in the complete kinematic plane because, as
pointed out before, for  $Q^2$ below 1 GeV$^2$, $W$ is already
approximately a function of only one variable, namely $x$,
and the data shows a smooth matching of the two
behaviors. Thus, $\tilde{\sigma}_{\gamma^*\mathrm{p}}$ should display
approximated geometric scaling for all values of $x$ and $Q^2$.

\begin{figure}
\begin{center}
\includegraphics*[width=.98\textwidth]{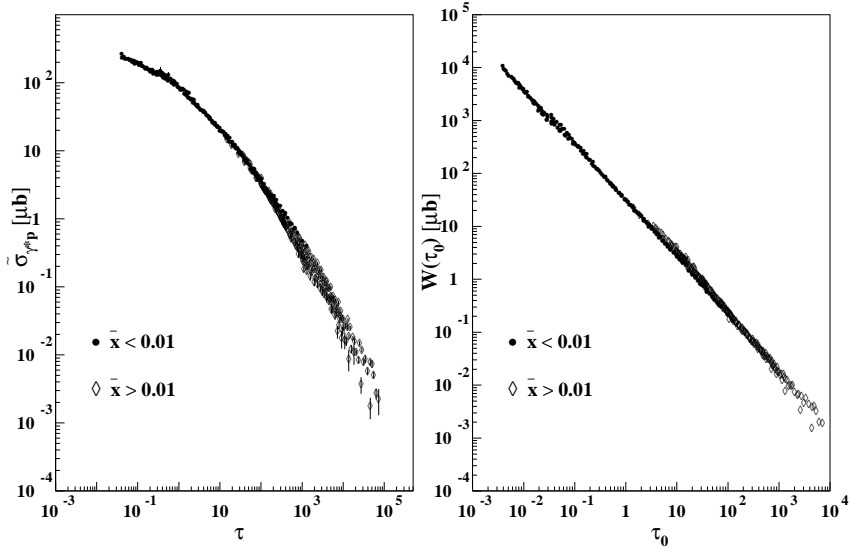}
\end{center}
\caption{The left panel shows the reduced $\gamma^*\mathrm{p}$ cross
section, $\tilde{\sigma}_{\gamma^*\mathrm{p}}$, as a function of the
scaling variable $\tau=Q^2/Q^2_s$~\cite{gbw} for all data
in~\cite{bcdms,h1data,zeus}.  The geometric scaling behavior is
clearly seen for all values of $\tau$. The right panel shows the
scaling combination of equation (\ref{eq_ngs}).  Data points in the
small $x$ region, $x<0.01$ are shown as full bullets to have a
comparison with previous studies~\cite{stasto}.}
\label{fig2} 
\end{figure}

The left panel of Figure \ref{fig2} shows that this is indeed the
case, while the right panel shows the scaling combination of equation
(\ref{eq_ngs}). Note that this latter case can not be considered
properly as scaling and that it is shown for illustration purposes only.

%%%%%%%%%%%%%%%%%%%%%%%%%%%%%%%%%%%%%%%%%%%%%%%%%%%%%%%%
%%%%%%%%%%%%%%%%%%%%%%%%%%%%%%%%%%%%%%%%%%%%%%%%%%%%%%%%

\begin{figure}
\begin{center}
\includegraphics*[width=.98\textwidth]{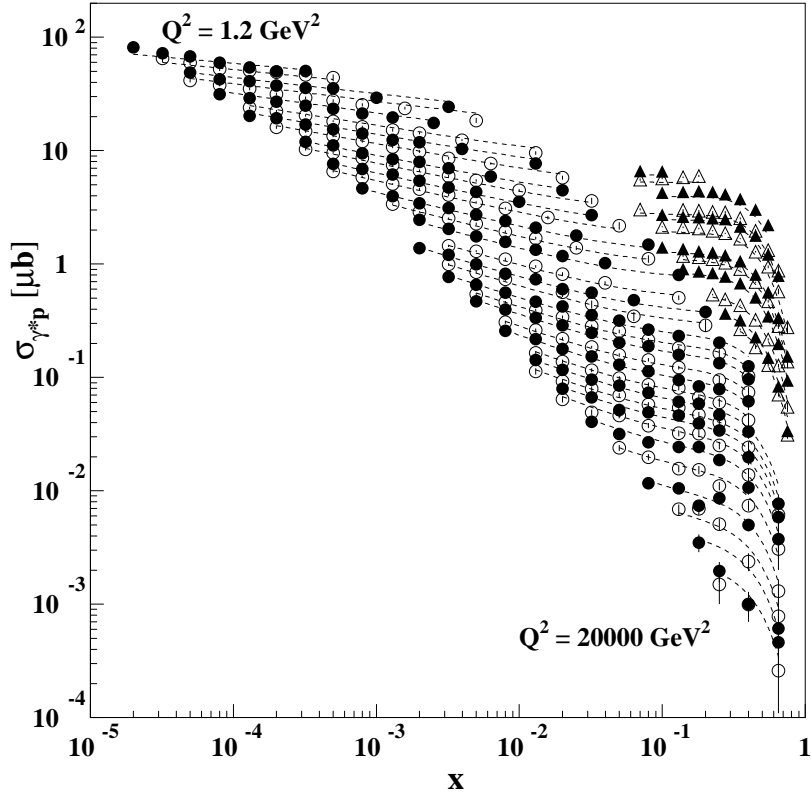}
\end{center}
\caption{The total $\gamma^*$p cross section,
$\sigma_{\gamma^*\mathrm{p}}$, is shown as a function of $x$ for
different fixed values of $Q^2$ going from $Q^2=1.2$ GeV$^2$ to
$Q^2=20000$ GeV$^2$. The bullets are the experimental data
points from HERA, while the triangles are from fixed target experiments.
The alternation of full and empty symbols is just to get a clear
display of data. The lines are the result of equation
(\ref{eq_all}). }
\label{fig4}
\end{figure}

The data span some six orders of magnitude in $x$, and
another six in $Q^2$.  With the addition of data with $x>0.01$ the
geometric scaling behavior is extended two orders of magnitude in
$\tau$.  It is quite interesting to compare Figure~\ref{fig2} with 
Figure~\ref{fig4}. The latter figure contains all data points
above $Q^2=1$ GeV$^2$ before the data collapse produced by the
transformation to the scaling variable $\tau$. The comparison of both
figures shows that the collapse of all data in a single line is not a
trivial fact.

{\bf Power laws, scaling and critical phenomena.}  It has to be
emphasized that the origin of scaling within this approach is the fact
that $W$ is very close to a generalized homogeneous function. This
fact is valid even at very large values of $x$, where one would not
necessarily expect saturation effects to be present. But it does not
exclude the possibility that the mechanism which gives rise to the
power law behavior of $W$ is also linked to saturation.

In this context it is interesting to note that scaling and its
relation to power laws has been widely discussed in relation to
critical phenomena. In particular it has been found that under some
conditions the presence of a renormalization group equation helps to
explain the appearance of power laws, of its associated scaling and
permits also to explain and numerically estimate the appearance of
scaling violations (see for example \cite{fisher,wilson} and
references therein).  

It is interesting to note that the power law behavior of the total
$\gamma^*\mathrm{p}$ cross section is generated by the branching
process embodied in QCD evolution equations which are in fact a type
of renormalization group equations. Also the case of saturation has
been cast, within the Color Glass Condensate approach \cite{mclerran},
in the form of renormalization group equations. It is clear then, that
the subject of finding a deeper understanding of the relation between
renormalization group equations and the emergence of power laws in
pQCD deserves further studies.

{\bf A parametrization of
$\mathbf{\sigma_{\gamma^*\mathrm{p}}}$ above $\mathbf{Q^2=1}$
GeV$^\mathbf{2}$.}  Note that as a consequence of the description of
$\sigma_{\gamma^*\mathrm{p}}$ given by equation (\ref{eq_wv}) we also
have a simple six parameter description of the total $\gamma^*$p cross
section for {\em all} $Q^2$ values above 1 GeV$^2$:

\begin{equation}
\begin{split}
\sigma_{\gamma^*\mathrm{p}}(\bar{x},Q^2)  = & 
\beta\left(\frac{Q^2}{Q^2_0}\right)^{-(1+\epsilon)}
\bar{x}^{-\alpha\log_{10}(Q^2/\Lambda^2)} \\
& \exp\left(-\frac{(\bar{x}-x_0)^2}{4\sigma^2}\right)
\mathrm{erf}\left(\frac{1-(x-x_0)}{2\sigma}\right).
\end{split}
\label{eq_all}
\end{equation}

%%%%%%%%%%%%%%%%%%%%%%%%%%%%%%%%%%%%%%%%%%%%%%%%%%%%%%%%
%%%%%%%%%%%%%%%%%%%%%%%%%%%%%%%%%%%%%%%%%%%%%%%%%%%%%%%%

Equation (\ref{eq_all}) is compared to  data in Figure \ref{fig4}
using the parameters obtained from the fit to Figure \ref{fig2}. For
these parameters the $\chi^2/dof$ obtained for $Q^2>1$ GeV$^2$ is
$\chi^2/dof=0.77$.

%%%%%%%%%%%%%%%%%%%%%%%%%%%%%%%%%%%%%%%%%%%%%%%%%%%%%%%%

\section{Summary and conclusions}
\label{s:summary} 

We have shown that $\sigma_{\gamma^*\mathrm{p}}$ can be separated as a
product of a power law function, $W$, carrying all the information on
its $Q^2$ evolution and a Gaussian--like distribution, $V$, which
depends only on $x$. As a consequence of this {\em Ansatz} we obtain a
six parameter description of all $\sigma_{\gamma^*\mathrm{p}}$ data
above $Q^2\approx 1$ GeV$^2$, where each parameter has a natural
physical interpretation. 

We define a reduced cross section,
$W\equiv\tilde{\sigma}_{\gamma^*\mathrm{p}}$, and find that it is very
close to a generalized homogeneous function over all the measured
kinematic plane. This property is found to be responsible for the
geometric scaling behavior of $\sigma_{\gamma^*\mathrm{p}}$ in the
small $x$ region and of $\tilde{\sigma}_{\gamma^*\mathrm{p}}$ in the
complete kinematic plane.  These results show that the emergence of
geometric scaling is not necessary related to saturation and open up
interesting possibilities for further studies of the relation between
evolution equations and the appearance of scaling behavior in QCD.

{\em Acknowledgments} This work has been partially supported by Conacyt
through grant 40073-F.

\end{document}